\documentclass[showpacs,preprintnumbers,amsmath,amssymb]{revtex4}

\usepackage[dvips]{color}
\usepackage{graphicx}

\newcommand{\al}{\alpha}

\newcommand{\De}{\Delta}

\newcommand{\eps}{\epsilon}

\newcommand{\La}{\Lambda}

\newcommand{\om}{\omega}
\newcommand{\Om}{\Omega}

\newcommand{\p}{\partial}
\newcommand{\<}{\langle} 
\renewcommand{\>}{\rangle} 
\newcommand{\txt}{\textstyle}
\newcommand{\dsp}{\displaystyle}
\newcommand\eqn[1]{(\ref{#1})}      
\newcommand\Eqn[1]{Eq.~(\ref{#1})}  
\newcommand{\beq}{\begin{equation}}
\newcommand{\eeq}{\end{equation}}
\newcommand{\ba}{\begin{array}}
\newcommand{\ea}{\end{array}}
\newcommand{\bea}{\begin{eqnarray}}
\newcommand{\eea}{\end{eqnarray}}
\newcommand{\bi}{\begin{itemize}}  
\newcommand{\ei}{\end{itemize}}
\newcommand{\ben}{\begin{enumerate}} 
\newcommand{\een}{\end{enumerate}}

\newcommand{\half} {{\txt \frac{1}{2}}}

\newcommand{\Det}{\mbox{Det}}

\newcommand{\hide}[1]{}

\newcommand{\delpp}{\Delta_{pp}}
\newcommand{\delnn}{\Delta_{nn}}
\newcommand{\delpn}{\Delta_{pn}}

\newcommand{\up}{\uparrow}
\newcommand{\dn}{\downarrow}
\newcommand{\MeV}{{\rm MeV}}

\newcounter{comcounter}

\usepackage[normalem]{ulem}  
\newcommand{\new}[1]{#1}

\begin{document}

\preprint{}

\title{Isospin asymmetry and type-I
\new{superconductivity}
in neutron star matter}

\author{Mark Alford$^1$}\author{Gerald Good$^1$}\author{Sanjay Reddy$^2$}%
\affiliation{$^1$Physics Department, Washington University,
St.~Louis, MO~63130, USA}
\affiliation{$^2$Theoretical Division, Los Alamos National Laboratory,
Los Alamos, NM 87545, USA}

\date{August 22, 2005}

\begin{abstract}
It has been argued by Buckley~et.~al.~\cite{zhitnitsky} that nuclear
matter is a type-I rather than a type-II superconductor. The suggested
mechanism is a strong interaction between neutron and proton Cooper
pairs, which arises from an assumed $U(2)$ symmetry of the effective
potential, which is supposed to originate in isospin symmetry of the
underlying nuclear interactions. To test this claim, we perform an
explicit mean-field calculation of the effective potential of the
Cooper pairs in a model with a simple four-point pairing
interaction. In the neutron star context, matter is very neutron rich
with less than 10\% protons, so there is no neutron-proton pairing.
We find that under these conditions our model shows no interaction
between proton Cooper pairs and neutron Cooper pairs at the mean-field
level. We estimate the leading contribution beyond mean field and find
that it is is small and attractive at weak coupling.
\end{abstract}


\maketitle

Recently, it has been suggested that nuclear matter in neutron stars
might be a type-I superconductor. The astrophysical evidence
is that certain neutron stars have long precession periods,
and it has been suggested that this means the proton superconductivity
cannot be type-II \cite{blink}, although that inference has been
contested \cite{Jones:2004xa,Sedrakian:2004yq}.
A theoretical argument for type-I superconductivity in 
neutron star cores was then presented by Buckley~et.~al. in 
Ref.~\cite{zhitnitsky}. These authors assumed that 
the effective potential for the Cooper pair fields
has a $U(2)$ symmetry under
rotation of proton and neutron Cooper pairs into
each other.
Specifically, the effective potential for the Cooper pair fields
was assumed to take the form
\beq
\label{eqn:badpotential}
\ba{c}
  V(|\De_{pp}|^2,|\De_{nn}|^2) \approx 
  U(|\De_{pp}|^2 + |\De_{nn}|^2) \\[1ex]
= -\mu_c(|\De_{pp}|^2 + |\De_{nn}|^2)
  + a (|\De_{pp}|^2 + |\De_{nn}|^2)^2 + \cdots
\ea
\eeq
The assumption that the potential is approximately a function of
$|\De_{pp}|^2 + |\De_{nn}|^2$ leads to $|\De_{pp}|^2|\De_{nn}|^2$
cross terms, which provide a
strong repulsive interaction between the neutron and proton
Cooper pair condensates. It is argued in Ref.~\cite{zhitnitsky}
that this leads to a long-range attraction between proton flux tubes,
i.e.~type-I superconductivity.

The assumed $U(2)$ symmetry was justified by invoking the isospin
symmetry of the underlying nuclear interaction. The authors of
Ref.~\cite{zhitnitsky} admit that isospin is severely broken by the
constraint of electrical neutrality, which,
\new{combined with beta-equilibration}, greatly suppresses the
proton Fermi momentum relative to the neutron Fermi momentum,
but they claim that this does not
affect the interaction between the $p$-$p$ and $n$-$n$
Cooper pairs.  This seems
implausible, since it is well known in the theory of
superconductivity that
the coefficients of the terms in the Landau-Ginzburg effective theory,
including the quartic coupling $a$,
depend strongly on the Fermi momenta of the underlying fermions
\cite{FetterWalecka}.
For example, for a one component Fermi gas, just below the BCS 
critical point, the effective Landau-Ginzburg potential is 
\beq
\label{eqn:landauginzburg}
V(|\De|^2) = -\mu |\De|^2 + a |\De|^4 + \cdots \\[1ex] \eeq where $\mu
\approx N_F (T_c-T)/T_c$, and $N_F$ is the density of states at the
Fermi surface, so for non-relativistic fermions $N_F\propto m p_F$
where $m$ is the mass and $p_F$ is the Fermi momentum.  The minimum of
the potential occurs at $\De_0=\sqrt{\mu/(2a)}$, so the quartic
coefficient is $a =\mu/(2\Delta_0^2)$. The binding energy of the
condensate is then $V(\De_0)=-\half N_F \De_0^2$, a well-known result
in the superconductivity literature \cite{Tinkham}.  We see that
$\mu$, $a$, and $V(\De_0)$ are all very sensitive to the Fermi momenta
of the underlying fermions.  Clearly for protons and neutrons in a
neutron star, which have very different Fermi momenta but similar
pairing condensates $\De$, this will not give an effective potential
of the form \eqn{eqn:badpotential}.

We now back up this argument with a concrete calculation.  We work
with a very simple microscopic model for 
nucleon-nucleon interactions, and analyze the
pairing using the mean-field approximation.
We take into account the
requirements of electrical neutrality \new{and equilibration
under the weak interactions}, which disfavor neutron-proton
pairing. At zero temperature the potential is not accurately
described by an low-order polynomial like \eqn{eqn:badpotential}, but we
can expand around the mean field and obtain the quartic
coupling between small fluctuations $\phi_{nn}$, $\phi_{pp}$,
\beq 
\ba{rl}
\Om(\phi_{pp},\phi_{nn}) =\cdots +
\al_{pp}\phi_{pp}^4 +
\al_{nn}\phi_{nn}^4 +
\al_{np}\phi_{pp}^2 \phi_{nn}^2+\cdots \,.  
\ea 
\eeq 
By explicitly calculating these couplings we find that they do not
obey the $U(2)$ symmetry assumed in
Ref.~\cite{zhitnitsky}. In fact, 
in the mean field approximation our result \eqn{mean_field_quartic_coeffs}
shows no interaction at all between the proton and neutron Cooper
pairs: $\al_{np}=0$.  
This appears to be a generic mean-field result, and does not
depend on any specific features of the pairing interaction.  Further,
and as expected from the preceding discussion, we find that $\al_{pp}
\ll \al_{nn}$.  We then discuss the lowest-order corrections beyond the
mean-field approximation by calculating the effective interaction
between neutron and proton Cooper pairs diagrammatically. We show that
this interaction is sub-leading in the coupling and is negligible in
weak coupling. 
At temperatures close to the critical temperature,
where Landau-Ginzburg theory can be used to analyze vortex structure,
we find that the interaction between the $pp$ condensate and the
$nn$ is weak and repulsive, not strong and attractive as \Eqn{eqn:badpotential}
would imply.

Our simple model of neutrons and protons is 
\new{based on the isospin-symmetric} Lagrangian
\beq
\label{lagrangian}
\ba{rcl}
 {\cal L} &=& {\cal{L}}_{\rm kinetic} + {\cal{L}}_{\rm int}\ , \\[2ex]

  {\cal{L}}_{\rm kinetic} &=& \dsp N_{\alpha a}^\dag \left(
    \frac{\p}{\p \tau} - \frac{\nabla^2}{2m} - \mu_a \right) N_{\alpha
    a}\ ,\\[2ex] {\cal{L}}_{\rm int} &=& \dsp -\frac{G}{2} \left(
    N_{\alpha a}^\dag N_{\alpha a} \right)^2\ .  
\ea 
\eeq 
The nucleon field $N_{a\al}$ has \new{isospin} index $a=n,p$ and 
spin index $\alpha=\up,\dn$. \new{Repeated indices are summed.
We immediately generalize the
interaction to allow different couplings for protons and
neutrons, and} Fierz-transform it into the pairing form.
\new{We assume that all pairing is in the rotationally
invariant $s$-wave channel, so keeping only those terms
we obtain}
\beq
\ba{rcl}
{\cal{L}}_{\rm int} &=& 
  -G_{pp}\, p_\up^\dag p_\dn^\dag p_\dn p_\up 
  -G_{nn}\, n_\up^\dag n_\dn^\dag n_\dn n_\up  \\[1ex]
 &-& \half G_{np}\,(p_\up^\dag
n_\dn^\dag + n_\up^\dag p_\dn^\dag)(n_\dn p_\up + p_\dn n_\up) \ .
\ea
\label{Lint}
\eeq
The isospin-symmetric case corresponds to $G_{pp}=G_{nn}=G_{np}$.
This interaction will lead to pairing of the fermions at their
Fermi surfaces, by the usual BCS mechanism. We can calculate the 
\new{thermodynamic potential}
of the paired state by a standard Hubbard-Stratonovich
transformation that introduces complex bosonic Cooper-pair fields
$\delnn, \delpp, $ and $\delpn$ (for a review, see \cite{stone}).  The
Lagrangian then has three parts: the kinetic term is as before, and
the others contain the Cooper pair fields: 
\beq
\label{lagrangian2}
\ba{rcl}
  {\cal{L}} &=& {\cal{L}}_{kinetic} + {\cal{L}}_{\Delta} + {\cal{L}}_F, \\[1ex]
  {\cal{L}}_{\Delta} &=&\dsp 
   |\delnn|^2/G_{nn} + |\delpp|^2/G_{pp} + |\delpn|^2/G_{np} , \\[2ex]
  {\cal{L}}_F &=& -\delnn^* n_\dn n_\up -\delpp^* p_\dn p_\up \\
 && -\dsp\frac{1}{\sqrt{2}}\delpn^* (p_\dn n_\up + n_\dn p_\up) + h.c.
\ea
\eeq
By adjusting the phases of the $n$ and $p$ fields we can choose
$\delnn$ and $\delpp$ to be real, leaving $\delpn$ complex.
\new{As is well known \cite{Delpnzero}, 
in neutron star matter the mismatch between
the proton and neutron Fermi surfaces completely suppresses
$n$-$p$ pairing, so $\delpn=0$ even at small isospin asymmetry, but 
for now we keep the $\delpn$ term.}
In the mean-field approximation, we neglect any space-time
variation in the $\Delta$ fields and set them equal to their vacuum
expectation values. We can then integrate out the fermions, giving us
\new{the volume density of the
thermodynamic potential or grand canonical potential
$\Om = E/V - \mu N/V = -p$.  In the rest of this paper
we will loosely refer to this
as the ``thermodynamic potential''.
We find}
\beq
  \Omega = {\cal{L}}_{\Delta} - 
    \int \frac{d\om}{2\pi} \int^\La\frac{d^3 k}{(2\pi)^3} \ln \Det M ,
\eeq 
where we have introduced an ultraviolet cutoff $\La$ on the
three-dimensional momentum integral.
\new{To obtain this expression, we wrote the inverse propagator in the 
Nambu-Gork'ov basis
$(p_\up,\, n_\up,\, p_\dn^\dag,\, n_\dn^\dag,\,
  p_\dn,\, n_\dn,\, p_\up^\dag,\, n_\dn^\dag)$,
and observed that it block-diagonalizes into two identical $4\times 4$ blocks,
}
\beq
   M = \left( \begin{array}{cccc} 
     -i\om + \eps_p & 0 & -\delpp & -\dsp\frac{\delpn}{\sqrt{2}} \\
     0 & -i\om + \eps_n & -\dsp\frac{\delpn}{\sqrt{2}} & -\delnn \\
     -\delpp & -\dsp\frac{\delpn^*}{\sqrt{2}} & -i\om -\eps_p & 0 \\
     -\dsp\frac{\delpn^*}{\sqrt{2}} & -\delnn & 0 & -i\om -\eps_n \\
                \end{array} \right).
\label{Mblock}
\eeq
\new{Note that when $\delpn=0$, $M$ further decomposes into $2\times 2$
blocks, for $(p_\up,\, p_\dn^\dag)$, $(n_\up,\, n_\dn^\dag)$ etc,
which describe the $s$-wave pairing of protons and neutrons respectively.}
In Eq.~\eqn{Mblock},
$\eps_p \equiv k^2/2m - \mu_p$ and $\eps_n \equiv k^2/2m - \mu_n$. 
The determinant is straightforward to compute and we obtain
\bea \Det M &=& (\om^2+\eps_n^2)(\om^2+\eps_p^2) + \delnn^2
  (\om^2+\eps_p^2) \notag \\ &+& \delpp^2 (\om^2+\eps_n^2) +
  |\delpn|^2 (\om^2 + \eps_p \eps_n) \notag \\ &+& (\delnn\delpp -
  \delpn^2/2)(\delnn\delpp - \delpn^{*2}/2) \eea
The requirement that the vacuum expectation values $\delpp, \delnn,$
and $\delpn$ minimize the \new{thermodynamic potential} 
is expressed in the three gap equations, 
\beq
\ba{r@{\,}c@{}l}
\dsp\frac{\delnn}{G_{nn}} &=& \dsp
  \int \!\!\frac{d\om d^3 k}{(2\pi)^4}
  \frac{\delnn(\om^2\!+\!\eps_p^2)
  \!+\!\delpp(\delpp\delnn\!-\!\mbox{Re}\delpn^2/2)}{\det M}  \\[2ex]
\dsp\frac{\delpp}{G_{pp}} &=& \dsp
  \int \!\!\frac{d\om d^3 k}{(2\pi)^4}
  \frac{\delpp(\om^2\!+\!\eps_n^2)
  \!+\!\delnn(\delpp \delnn\!-\!\mbox{Re}\delpn^2/2)} {\det M}  \\[2ex]
\dsp\frac{\delpn}{G_{np}} &=& \dsp
  \int \!\!\frac{d\om d^3 k}{(2\pi)^4}
  \frac{\delpn (\om^2\!+\!\eps_p \eps_n)
  \!-\!\delpn^*(\delpp \delnn\!-\!\delpn^2/2)}{\det M} 
\ea
\eeq
%
When $\De_{pn}=0$, \new{as in neutron star matter \cite{Delpnzero}}
these equations decouple into independent
gap equations for $\De_{pp}$ and $\De_{nn}$.

Because our model interaction is so simple, it predicts pairing gaps
that rise with the density of states near the Fermi surface, so $\De$
rises with $p_F$. This means that it does not produce realistic
neutral nuclear matter, in which the interaction is isospin-symmetric,
and $p_{Fp}\ll p_{Fn}$ (from electrical neutrality) but nevertheless
$\De_{pp}\approx\De_{nn}$. In the real world, this happens because the
nuclear interaction at short distance is repulsive, so even though
$p_{Fn}\gg p_{Fp}$, the neutrons end up with a similar pairing gap to
the protons. In our simple model, which is strictly valid only at low
density, the four-Fermion coupling only encodes the attractive part of
the interaction through the s-wave scattering length. We will choose
$G_{pp}$ to be larger than $G_{nn}$ so that in the neutral system the
protons and neutrons have the same pairing gap, as in real nuclear
matter. It will turn out that our essential conclusion, that there is
no $U(2)$ symmetry of the effective potential for the Cooper pair
fields, holds irrespective of whether or not the the couplings $G_{pp}$
and $G_{nn}$ are equal.  It is then reasonable to guess that $G_{np}$
should have a value somewhere between $G_{nn}$ and $G_{pp}$. To be
specific, we shall employ typical values $\mu_n \sim $ 60 MeV
($k_{Fn}\sim 335~\MeV$) and $\mu_p \sim$ 8 MeV ($k_{Fp}\sim 123~\MeV$)
and a momentum cut-off $\La=750~\MeV$. For these parameters the
four-fermion couplings that gives $\De_{nn}\sim 1$ MeV and $\De_{pp}
\sim 1 $ MeV are $G_{nn} \sim 1\times 10^{-5}~\MeV^2$ and $G_{pp} \sim
2\times 10^{-5}~\MeV^2$, respectively.

\new{Since $\De_{np}=0$ in beta-equilibrated neutral nuclear matter,
we can write the mean-field thermodynamic potential}
\begin{eqnarray}
\Omega &=& \frac{\delnn^2}{G_{nn}} + \frac{\delpp^2}{G_{pp}} 
- \int \frac{d\om}{2\pi} \int^\La \frac{d^3 k}{(2\pi)^3}
 \Bigl( \ln (\om^2 + \eps_n^2 + \delnn^2) 
 + \ln (\om^2 + \eps_p^2 + \delpp^2) \Bigr) \\
 &=& \frac{\delnn^2}{G_{nn}} - \int^\La\frac{d^3 k}{(2\pi)^3} \Bigl( \sqrt{\eps_n^2 + \delnn^2}-\eps_n\Bigr) + \frac{\delpp^2}{G_{pp}} - \int^\La \frac{d^3 k}{(2\pi)^3} \Bigl(\sqrt{\eps_n^2 + \delnn^2}-\eps_p \Bigr)
\label{eqn:free-energy}
\end{eqnarray}
We see that this is equal to the sum of \new{the thermodynamic potentials}
for two species of Cooper pair bosons
that do not interact with each other. There are no cross-terms between
$\De_{pp}$ and $\De_{nn}$.
The effective potential does not take the form (\ref{eqn:badpotential}).
This result did not
depend on the specific form of the interaction.
It simply arises from the fact
that \new{beta-equilibrium and}
electrical neutrality require $p_{Fp}\ll p_{Fn}$, and
Cooper pairing is
suppressed between species with very different Fermi momenta.

\begin{figure}[floatfix]
\includegraphics[width=0.3\textwidth]{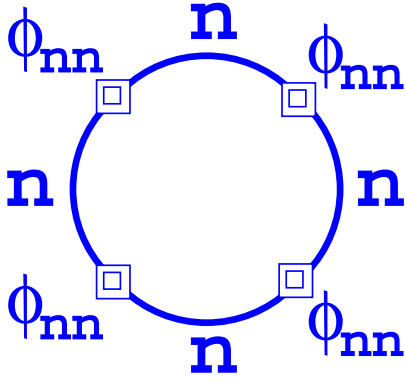}
\caption{
 The lowest-order mean-field contribution to \new{$\al_{nn}$, 
the coefficient of the $\phi_{nn}^4$ term in the effective potential,
which describes} scattering between \new{low-momentum} fluctuations 
$\phi_{nn}$ in the $nn$ condensate. 
\new{The thick lines are Nambu-Gork'ov
neutron propagators.
Each double-square vertex is an insertion of
the $\phi_{nn}$ operator.} There is a similar diagram for $\al_{pp}$.
}
\label{fig:nnnn_fc_diag}
\end{figure}

We may use Eq.~\eqn{eqn:free-energy} to investigate the nature of the
quartic terms that describe the coupling between fluctuations in the
Cooper-pair densities. Fluctuations in the pairing field $\phi_{nn},
\phi_{pp}$ are defined through the following substitutions in
Eq.~\eqn{eqn:free-energy}: 
$\Delta_{nn}\rightarrow \tilde{\Delta}_{nn}+\phi_{nn}$ 
and $\Delta_{pp}\rightarrow \tilde{\Delta}_{pp}+ \phi_{pp}$ 
where $\tilde{\Delta}_{nn}$ and $\tilde{\Delta}_{pp}$
are the ground state expectation values that satisfy the gap
equations. For small fluctuations an expansion of the 
\new{thermodynamic potential}  
about the mean field ground state is well motivated. 
Retaining only the quartic terms  
\beq 
\ba{rl}
\Om(\phi_{pp},\phi_{nn}) =\cdots +
\al_{pp}\phi_{pp}^4 +
\al_{nn}\phi_{nn}^4 +
\al_{np}\phi_{pp}^2 \phi_{nn}^2+\cdots \,.  
\ea 
\label{tzmftexp}
\eeq 
From the preceding discussion it is clear that in the mean field
approximation there are no cross-terms. The
coefficients $ \al_{nn}$ and $\al_{pp}$ are non zero and depend in
general on the chemical potentials, $\tilde{\Delta}_{nn}$ and
$\tilde{\Delta}_{pp}$. Explicitly, by Taylor expanding
Eq.~\eqn{eqn:free-energy} we find 
\beq
\ba{rcl}
\al_{np}&=&0 \, , \\[1ex]
\al_{nn}&=& \dsp \frac{1}{8} \int\frac{d^3k}{(2 \pi)^3}~\frac{1}{E_n^3}
  -\frac{6 \tilde{\Delta}_{nn}^2}{E_n^5}
  +\frac{5 \tilde{\Delta}_{nn}^4}{E_n^7} \,, \\[2ex]
\al_{pp}&=& \dsp \frac{1}{8} \int\frac{d^3k}{(2\pi)^3}~\frac{1}{E_p^3} 
  -\frac{6 \tilde{\Delta}_{pp}^2}{E_p^5} 
  +\frac{5 \tilde{\Delta}_{pp}^4}{E_p^7} \,, 
\ea
\label{mean_field_quartic_coeffs}
\eeq
where $E_{n}=\sqrt{\epsilon_n^2+\tilde{\Delta}_{nn}^2}$ and
$E_{p}=\sqrt{\epsilon_p^2+\tilde{\Delta}_{pp}^2}$. These contributions
can also be calculated in a diagrammatic approach using
the Nambu-Gorkov Greens functions \cite{Schrieffer}. The diagram
for $\al_{nn}$ is shown in Fig.~\ref{fig:nnnn_fc_diag}, and
there is an analogous one for $\al_{pp}$. We find

\begin{equation} 
\alpha_{nn} = \frac{1}{4} \int~\frac{d^3k}{(2 \pi)^3}~kT\sum_{s}~{\rm Tr}~\left[{\cal G}_n(k,i \omega_s)\tau {\cal G}_n(k,i \omega_s)\tau {\cal G}_n(k,i \omega_s)\tau {\cal G}_n(k,i \omega_s)\tau\right] \,,
\end{equation}
where the finite temperature Nambu-Gorkov Green's function for the neutron 
superfluid,
\new{expressed in the $(n_\up, n_\dn^\dag)$ basis (see text after
Eq.~\eqn{Mblock})}, is
\beq
\label{NGG}
{\cal G}_n(k,i\omega_s)=\frac{1}{(i\om_s)^2-E_n^2(p)}~
   \left( \begin{array}{cc} 
     -i\om_s + \eps_n & -\Delta_{nn} \\
     -\Delta_{nn} & -i\om_s - \eps_n \\
         \end{array} \right)\,,
\eeq
and insertions of the fluctuating pairing field are
\beq
 \tau = \left(
  \begin{array}{cc} 
     0 & 1 \\
     1 & 0 \\
  \end{array} \right)\, .
\eeq
The Matsubara frequency is $\om_s=(2s+1)\pi kT$. In evaluating these
diagrams we ignore any momentum transfer since we are interested only
in the low momentum fluctuations.  We have explicitly checked that
this gives the same result as \Eqn{mean_field_quartic_coeffs}.
At zero temperature and when $\tilde{\Delta}/ \mu$
is small we obtain the following analytic expressions
\begin{eqnarray}
\al_{nn}&=& -\frac{M k_{Fn}}{24 \pi^2 \tilde{\Delta}_{nn}^2}~
\left(1 + {\rm O}[\frac{\tilde{\Delta}_{nn}^2}{\mu_n^2}]\right)\,, \\
\al_{pp}&=& -\frac{M k_{Fp}}{24 \pi^2 \tilde{\Delta}_{pp}^2}
\left(1 + {\rm O}[\frac{\tilde{\Delta}_{pp}^2}{\mu_p^2}]\right)
\,.
\end{eqnarray}
We see that the couplings are proportional to the corresponding Fermi
momenta, $\al_{nn}\propto k_{Fn}$ and $\al_{pp}\propto k_{Fp}$, so the
scattering of proton condensate fluctuations is much weaker than that
of neutron condensate fluctuations, indicating a strong breaking of
any symmetry that rotates proton Cooper pairs into neutron Cooper
pairs. Further, the sign of $\al_{nn}$ and $\al_{pp}$ is
negative. However, since we are Taylor expanding about the global
minimum of the \new{thermodynamic potential}, 
there are lower and higher order terms (not
explicitly written in Eq.~(\ref{tzmftexp})) that ensure that the system is
stable with respect to both small and large fluctuations. This is in
contrast to the usual Landau- Ginzburg analysis just below $T_c$,
where a Taylor expansion about the normal state is characterized by a
negative quadratic coefficient and a positive quartic term.

\begin{figure}[floatfix]
\includegraphics[width=0.4\textwidth]{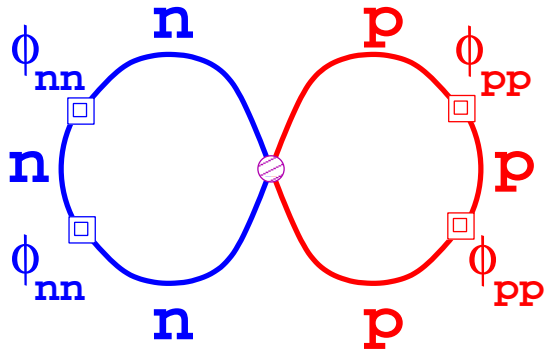}
\caption{ The lowest-order contributions beyond mean field
to \new{$\al_{np}$, the
coefficient of the $\phi_{nn}^2\phi_{pp}^2$ term in the effective potential,
which describes} scattering between \new{low-momentum} fluctuations 
$\phi_{nn}$ and $\phi_{pp}$ in the $nn$ and $pp$ condensates. 
\new{
The thick lines are Nambu-Gork'ov
neutron or proton propagators.
The double square vertices are insertions of the
$\phi_{nn}$ and $\phi_{pp}$ operators.}
The hatched vertex is the 
fundamental $n^\dag n p^\dag p$ interaction in the Lagrangian.
}
\label{fig:nnpp_fc_diag}
\end{figure}
We now discuss corrections to Eq.~\eqn{eqn:free-energy} beyond
the mean field approximation. The leading order diagram that
contributes to the scattering between neutron and proton fluctuations
is shown in Fig.~\ref{fig:nnpp_fc_diag}. 
The important point is that this diagram involves
the fundamental four-fermion
neutron-proton interaction $G_{np}$. Evaluating the diagram, we find
that the leading-order beyond-mean-field contribution to the effective
four-point interaction between the proton and neutron condensate
fluctuations is 
\begin{eqnarray}
 \al_{np} &=& \frac{1}{4}~G_{np}\int \frac{d^4k}{(2\pi)^4}~{\rm Tr}\left[{\cal G}_n(k,i\om)\tau {\cal G}_n(k,i\om)\tau {\cal G}_n(k,i\om)\tau_3\right]\nonumber \\
 &~&\times \int \frac{d^4k}{(2\pi)^4}
 {\rm Tr}\left[{\cal G}_p(k,i\om)\tau {\cal G}_p(k,i\om) \tau {\cal G}_p(k,i\om)\tau_3\right] \\
 &=& -~G_{np}~\frac{k_{Fn}^3 k_{Fp}^3}{64\pi^4 \mu_n^2 \mu_p^2}~f(\frac{\delnn}{\mu_n})~f(\frac{\delpp}{\mu_p})\,,
\label{anp}
\end{eqnarray}
where $ {\cal G}_n(k,i\om)$ and $ {\cal G}_p(k,i\om)$ are the
Nambu-Gorkov Green's functions for the neutrons and protons
respectively, and $\tau_3={\rm diag}(1,-1) $ is the diagonal Pauli
matrix \cite{Schrieffer}. The dimensionless function $f$ is defined by
the integral
\beq
f(\delta)=\int_{-1}^{\infty}~ dx~x~\sqrt{x+1}~\frac{x^2-2\delta^2 }{
    (\delta^2 + x^2)^{5/2} }\,.
\eeq
For $\delta \ll 1$ we obtain the following analytic relation
\beq
f(\delta)=-1+\log\left[\frac{8}{\delta}\right] + {\cal O}[\delta^2]\,.
\eeq

To see whether beyond-mean-field corrections can raise $\al_{np}$ to
a value comparable to $\al_{nn}$ or $\al_{pp}$ we follow Ref.~\cite{zhitnitsky}
in defining an asymmetry parameter 
\beq
\eps = (\al_{nn} \al_{pp} - \al_{np}^2)/(\al_{nn} \al_{pp})\ .
\eeq
(Ref.~\cite{zhitnitsky} expresses $\eps$ in terms of the quartic
couplings $a_{pp}$, $a_{nn}$, $a_{np}$ for the expansion around
$\De=0$, but it should be equally valid to expand around the mean field,
which is the minimum of the \new{thermodynamic potential}.)
Ref.~\cite{zhitnitsky} concluded that type-I superconductivity requires
$\eps \lesssim 1/20$. As we have seen, in the mean-field approximation
$\eps=1$. From \Eqn{anp} we find that the beyond-mean-field corrections
to $\eps$ are negligible by several orders of magnitude.
$\eps$ is always within $10^{-4}$ of unity. 


All our calculations so far have been at zero temperature, in which
case the expansion of the potential around the mean-field ground state
is only valid for small deviations from that state. This makes it impossible
to discuss the expected structure of the vortices, in which the pairing
fields vary from zero to their vacuum values.
In order to be able to say anything about the vortices, we have to
work at temperatures close to the critical temperature $T_c$, where
a traditional Landau-Ginzburg analysis is possible, using
an expansion
around the zero-mean-field state with only the quadratic and quartic terms,
\begin{equation} 
\Omega(T\sim T_c;{\phi_{nn},\phi_{pp}}) =   -\mu^{c}_{nn} \phi_{nn}^2  -\mu^{c}_{pp}\phi_{pp}^2 + a^{c}_{pp}\phi_{pp}^4 + a^{c}_{nn}\phi_{nn}^4 + a^{c}_{np}\phi_{pp}^2 \phi_{nn}^2+\cdots\,.
\label{lgtc}
\end{equation}
We shall now perform such an analysis.
We assume that the critical temperatures of the neutrons
and protons are not vastly different and we perturb about the Fermi gas
ground state. The diagrammatic method described earlier permits us to
calculate the coefficients in Eq.~(\ref{lgtc}). Since we are perturbing
about the normal state, we use the Green's function defined in
Eq.~(\ref{NGG}) but with $\Delta_{nn}=\Delta_{pp}=0$. The quadratic
coefficient
\begin{eqnarray} 
\mu^{c}_{nn}&=&\frac{1}{G_{nn}} + \Pi_{nn} \,, ~
{\rm where} \quad \Pi_{nn} = \frac{1}{2} \int \frac{d^3k}{(2\pi)^3}~kT\sum_{s}~{\rm Tr}\left[{\cal G}^0_n(k,i\om_s)\tau {\cal G}^0_n(k,i\om_s)\tau\right]\,,\\
\mu^{c}_{pp}&=&\frac{1}{G_{pp}} + \Pi_{pp} \,,~
{\rm where} \quad \Pi_{pp} = \frac{1}{2} \int \frac{d^3k}{(2\pi)^3}~kT\sum_{s}~{\rm Tr}\left[{\cal G}^0_p(k,i\om_s)\tau {\cal G}^0_p(k,i\om_s)\tau\right]\,,
\end{eqnarray}
where ${\cal G}^0_n(k,i\om_s)=\lim_{\Delta_n\rightarrow0}{\cal G}_n(k,i\om_s)$ 
and  ${\cal G}^0_p(k,i\om_s)=\lim_{\Delta_p\rightarrow0}{\cal G}_p(k,i\om_s)$. 
An explicit calculation yields
\begin{eqnarray}
\mu^{c}_{nn}&=&\frac{1}{G_{nn}} - \frac{(2m kT)^{3/2}}{8 \pi^2~ kT} J_1(\frac{\mu_n}{kT})\,,\\
\mu^{c}_{pp}&=&\frac{1}{G_{pp}} - \frac{(2m kT)^{3/2}}{8 \pi^2~kT} J_1(\frac{\mu_p}{kT})\,,
\end{eqnarray}
where the dimensionless integral
\begin{equation} 
J_1(\gamma) = \int_{0}^{\tilde{\Lambda}}~dx ~\frac{\sqrt{x}}{x-\gamma}~\tanh{\left[\frac{x-\gamma}{2}\right]}\,,
\end{equation} 
and $\tilde{\Lambda}=\Lambda^2/(2mkT)$. 
The critical temperature $T_c$ for neutron and protons is defined by
the relation $\mu^{c}_{nn}=0$ and $\mu^{c}_{pp}=0$, respectively. As
expected, for $T>T_c$ we obtain $\mu^{c}_{nn},\mu^{c}_{pp}>0$ while
$\mu^{c}_{nn},\mu^{c}_{pp}<0$ for $T<T_c$.  The quartic coefficient is
\begin{eqnarray} 
a^c_{nn}=\frac{1}{4}~ \int \frac{d^3k}{(2\pi)^3}~kT\sum_{s}~{\rm Tr}\left[{\cal G}^0_n(k,i\om_s)\tau {\cal G}^0_n(k,i\om_s)\tau {\cal G}^0_n(k,i\om_s)\tau {\cal G}^0_n(k,i\om_s)\tau  \right]\,,
\end{eqnarray}
and $a^c_{pp}$ is given by the same relation but with the  ${\cal G}^0_p(k,i\om_s)$ replacing   ${\cal G}^0_n(k,i\om_s)$. The Matsubara sum is easily evaluated and we obtain 
\begin{equation} 
a^c_{nn}=\frac{(2m~kT)^{3/2}}{64 \pi^2~(kT)^3}~J_3(\frac{\mu_n}{kT})\,,
\end{equation}
where 
\begin{equation} 
J_3(\gamma) = \int_{0}^{\tilde{\Lambda}}~dx ~\frac{\sqrt{x}}{(x-\gamma)^3}~\frac{\sinh{[x-\gamma]}-(x-\gamma)}{(1+\cosh{[x-\gamma]})}\,.
\end{equation} 
The coefficient that couples between neutron and proton pairing fields 
is given by 
\begin{equation} 
a^c_{np}=-~\frac{G_{np}~m^{3}}{32 \pi^4~(kT)}~J_2(\frac{\mu_n}{kT}) \times J_2(\frac{\mu_p}{kT})\,,
\end{equation}
where 
\begin{equation} 
J_2(\gamma) = \int_{0}^{\tilde{\Lambda}}~dx ~\frac{\sqrt{x}}{(x-\gamma)^2}~\frac{\sinh{[x-\gamma]}-(x-\gamma)}{(1+\cosh{[x-\gamma]})}\,.
\end{equation} 
Thus in the vicinity of the critical temperature the asymmetry
parameter $\epsilon$ can be estimated using $a^c_{nn}$, $a^c_{pp}$ and
$a^c_{np}$ calculated above.  For our choice of parameters  we find that $T_c\simeq  0.6~\MeV$. 
For $T=0.4~\MeV$ the coefficients of the effective potential are
$\mu^c_{nn}=5520~\MeV^2$,
$\mu^c_{pp}=2010~\MeV^2$, $a^c_{nn}=2627$, $a^c_{pp}=968$ and
$a^c_{np}=-3 (G_{np}/G_{nn})$. As before, the asymmetry parameter
$\epsilon \simeq 1$ since $a_{np} \ll a_{nn}$ and $a_{np} \ll a_{pp}$.
The fact that $a^c_{np}$ is negative implies that
the neutron(proton) superfluid density will decrease in the inner core of the
proton(neutron) vortex. 

Our conclusion is that in neutral nuclear matter, the disparity
between the neutron and proton Fermi momenta 
provides a strong explicit breaking
of the $U(2)$ symmetry posited in Ref.~\cite{zhitnitsky}. This
breaking is far too strong to allow the proposed mechanism for type-I
superconductivity to operate. Although a calculation of the vortex structure is beyond the scope of this work, our leading order calculation of $a^c_{np}$ 
provides a means to estimate the strength of the effective interaction between
neutron and proton condensate vortices, at temperatures close to $T_c$. We note however that due to the
non-perturbative nature of the interaction between nucleons we cannot
exclude the possibility of a strong coupling between the neutron and
proton superfluids. In our simple  model the attractive interaction between neutrons 
and protons directly leads to a negative $a^c_{np}$ leading to a depletion of the neutron superfluid in the core of the proton vortex. This result is robust as long as the effective interaction  between neutrons and protons is attractive and contradicts the predictions in Ref.~\cite{zhitnitsky} where the neutron superfluid density increased inside the proton vortex. 

\vspace{3ex}
\begin{acknowledgments}
We acknowledge helpful conversations with T. Battacharya, M. Metlitski,
G. Rupak, and A. Zhitnitsky.
The work of
MGA and GG is supported by the U.S. Department of Energy under 
grant number DE-FG01-91ER40628. The work of SR is supported
by the U.S. Department of Energy under
grant number W-7405-ENG-36.
\end{acknowledgments}

\end{document}